\documentclass[preprint]{elsarticle}
\usepackage{amssymb}
\usepackage{graphicx}

\journal{Physics Letters B}
\begin{document}
\begin{frontmatter}
\title{Some properties of the ``String gas'' with the equation of state $p = -\frac13 \rho$}
\author{Alexander Y. Kamenshchik\corref{cor1}}
\address{Dipartimento di Fisica and INFN, Via Irnerio 46,40126 Bologna,
Italy\\
L.D. Landau Institute for Theoretical Physics of the Russian
Academy of Sciences, Kosygin str. 2, 119334 Moscow, Russia}\ead
{Alexander.Kamenshchik@bo.infn.it}
\author{Isaak M. Khalatnikov}
\address{L.D. Landau Institute for Theoretical Physics of the Russian
Academy of Sciences, Kosygin str. 2, 119334 Moscow, Russia}\ead{khalat@itp.ac.ru}
\cortext[]{Corresponding author}
\begin{abstract}
We show that the string gas - a perfect fluid with the equation of state $p = -\frac13 \rho$ possesses rather interesing properties.
In Friedmann universes its presence can can change the observable topology 
of the space; in the spherically symmetric spacetimes it produces rather bizzare geometries and in a way its influence on the 
rotation curves mimics the dark matter effects.  
\end{abstract}
\begin{keyword} 
dark energy, dark matter, Tolman-Oppenheimer-Volkoff equations
\PACS 98.80.Jk\sep 04.20.Jb
\end{keyword}
\end{frontmatter}

\section{Introduction}
The discovery of the cosmic acceleration \cite{cosmic} and the search for the  energy \cite{dark}, responsible for this phenomenon have stimulated interest for rather exotic types of matter. The present letter is devoted to some curious properties 
of the perfect fluid often called `` string gas'' or ``gas of strings''. The equation of state for this gas is
\begin{equation}
p = -\frac13 \rho,
\label{eq-of-state}
\end{equation}
where $p$ is the pressure and $\rho$ is the energy density. 
One can say that the string gas can be classified as a kind of perfect fluid, which has a rather natural geometrical interpretation. Indeed, the presence of this matter in the universe is equivalent to the existence of the spatial curvature.
We shall dwell on this point in the second section of the letter. Here we would like to add that between other ``geometrical'' species  of matter one can mention the cosmological constant with the equation of state $p = -\rho$ and the stiff matter 
with the equation of state $p = \rho$, whose presence is equivalent to the presence of some anisotropy in the spatially homogeneous universe \cite{Bel-Khal,Khal-Kam}. 

As is well known the condition of the accelerated expansion of the universe is $p < -\frac13 \rho$. Then the so called 
strong energy dominance condition $\rho + 3p > 0$ arises rather often in different contests. Thus, the study of properties 
of such a ``boundary'' case as a gas with the equation of state (\ref{eq-of-state}) is of certain interest. The structure of our letter is the following: in the second section we consider the Friedmann universes and show how the presence of the string gas can 
change the observable curvature of the universe; the third section is devoted to study the solutions of the Tolman-Oppenheimer-Volkoff equations in the presence of the string gas; the fourth section is devoted to the 
motion of point-like objects around the spherically symmetric object surrounded by the string gas (rotation curves);
The last section contains some concluding remarks. 

\section{Friedmann universe}
Let us consider the Friedmann universe filled with some mixture of perfect fluids with equations of state $p_i = w_i\rho_i$.
We suppose that one of these fluids is the string gas.
The metric is as usual 
\begin{equation}
ds^2 = dt^2 -a^2(t)\left(\frac{dr^2}{1-kr^2} + r^2(d\theta^2 + \sin^2\theta d\phi^2)\right),
\label{Fried}
\end{equation}
where $k =-1,0$ or $+1$ for open, flat and closed universes respectively. 
The first Friedmann equation is 
\begin{equation}
\frac{\dot{a}^2}{a^2}+\frac{k}{a^2} = \sum_{i=1}^{N} \frac{C_i}{a^{3(1+w_i)}}.
\label{Fried1} 
\end{equation}
It is obvious that for the case string gas equation of state the corresponding term in the right-hand side of Eq. (\ref{Fried1}) behaves 
like $\frac{C{s}}{a^2}$ i.e. it has the behaviour coinciding with that of the curvature term $\frac{k}{a^2}$ and can in some sense mimic it.
Indeed, let us consider a simple example: let us consider consider a closed Friedmann universe, containing some amount of the string gas.
Shifting the string gas energy density from the right-hand side of Eq. (\ref{Fried1}) to its left-hand side we obtain
\begin{equation}
\frac{\dot{a}^2}{a^2}+\frac{1-C_s}{a^2} = \rho,
\label{Fried2}
\end{equation}
where $\rho$ is the total energy density of all perfect fluids, excluding the string gas. If $C_s = 1$ then the equation (\ref{Fried2}) is 
undistinguishable from the Friedmann equation for the flat Friedmann universe in the absence of the string gas. If $C_s > 1$ the situation 
is even more curious - after the suitable rescaling of the cosmological radius $a$ (namely, intoducing $\tilde{a} \equiv a/\sqrt{C_s-1}$) 
we can come to the Friedmann equation for an open universe. Thus, the presence of the string gas can change the observable spatial topology of a universe. 

\section{Tolman-Oppenheimer-Volkoff equations in the presence of the string gas}
After the discovery of cosmic acceleration the study of such a classical topic as a spherically symmetric solutions of the Einstein equaitons,
namely - the study of the Tolman-Oppenheimer-Volokoff equation have acquired a new impact. The search for  such solutions in a spacetime filled 
with some kind of dark energy revealed a new and interesing geometries.
Here we consider the spherically symmetric static star-like object immersed into a universe,
filled with the string gas.
We shall look for a static spherically symmetric metric of the form:
\begin{equation}
ds^2 = e^{\nu(r)}dt^2 - e^{\mu(r)}dr^2 -r^2(d\theta^2 + \sin^2\theta d\phi^2).
\label{metricTOV}
 \end{equation}
The time-time radius-radius components of the Einstein equations are
\begin{equation}
e^{-\mu}\left(\frac{1}{r}\frac{d\mu}{dr}-\frac{1}{r^2}\right)+\frac{1}{r^2} = 8\pi \rho,
\label{Einstein0}
\end{equation}
\begin{equation}
e^{_\mu}\left(\frac{1}{r}\frac{d\nu}{dr}+\frac{1}{r^2}\right)-\frac{1}{r^2} = 8\pi p.
\label{Einsteinr}
\end{equation}
The energy-momentum conservation equation is
\begin{equation}
\frac{dp}{dr} = -\frac{d\nu}{dr}\frac{\rho+p}{2}.
\label{conserv}
\end{equation}
Solving Eq. (\ref{Einstein0}) with the boundary condition $e^{-\mu(0)} = 1$ gives
\begin{equation}
e^{-\mu} = \left(1-\frac{2M}{r}\right),
\label{mu}
\end{equation}
where $M(r) = 4\pi \int _0^r dr r^2 \rho(r)$. This is equivalent to
\begin{equation}
\frac{dM}{dr} = 4\pi r^2 \rho,\ M(0) = 0.
\label{massTOV}
\end{equation}

Equations (\ref{Einsteinr})--(\ref{mu}) give rise to the Tolman-Oppenheimer-Volkoff equations \cite{TOV1,TOV2}
\begin{equation}
\frac{dp}{dr} = -\frac{(\rho+p)(M+4\pi r^3p)}{r(r-2M)}.
\label{TOV}
\end{equation}
Substituting instead of the pressure $p$ the energy density taken from Eq. (\ref{eq-of-state}) we come to the following
equation
\begin{equation}
\frac{d\rho}{dr} = \frac{2\rho(M-\frac{4\pi r^3}{3}\rho)}{r(r-2M)}.
\label{TOV1}
\end{equation}
Thus, we shall study the system of two equations (\ref{massTOV}) and (\ref{TOV1}) for two variables $\rho$ and $M$.
Similar analysis for the Tolman-Oppenheimer-Volkoff equations in the presence of the Chaplygin gas and the generalized 
Chaplygin gas \cite{Chap} was done in \cite{we-TOV} and \cite{we-TOV1}.
Let us suppose that at the center of the universe there is a massive spherically symmetric object ``star'' with the radius
$r_b$ and the mass $M(r_b)$. The mass and the radius satisfy the ``absence of horizon'' relation
\begin{equation}
M(r_b) < \frac{r_b}{2}.
\label{hor}
\end{equation}
The star is immersed in the string gas, whose  energy density at the boundary of the star is  $\rho(r_b)$. We shall consider the situation when this initial energy density of the string gas is much less than the average density of the star, i.e.
\begin{equation}
\rho(r_b) \ll \frac{M(r_b)}{\frac43\pi r_b^3}.
\label{initial}
\end{equation}
In this case the right-hand side of Eq. (\ref{TOV1}) is positive and the energy density $\rho$ is growing with the radius $r$.
That means that the mass $M$ is growing faster than $r^3$, the factor $(r-2M)$ is diminishing and at some value $r_0$ this
factor disappear:
\begin{equation}
r_0 = 2M(r_0).
\label{Schwarz}
\end{equation}
At the same time the factor $(M-\frac{4\pi r^3}{3}\rho)$ in the numerator of the right-hand side of Eq. (\ref{TOV1}) is also decreasing. Indeed,using the equation (\ref{massTOV}) we see that
\begin{equation}
\frac{d(M-\frac{4\pi r^3}{3}\rho)}{dr} = -\frac{4\pi r^3}{3}\frac{d\rho}{dr} < 0.
\label{drho}
\end{equation}
The point is that the expressions $(r-2M)$ and $(M-\frac{4\pi r^3}{3}\rho)$ vanish at the same value of the radius $r_0$.
To prove this statement let us rewrite Eq. (\ref{TOV1}) in the following form
\begin{equation}
\frac{d\rho}{\rho - \frac{M}{\frac{4\pi r^3}{3}}} = - \frac{8\pi r^2 dr}{r-2M}.
\label{TOV2}
\end{equation}
Thus, if the factor $(r-2M)$ tends to zero while $\rho - \frac{M}{\frac{4\pi r^3}{3}}$ is finite the left hand side
of Eq. (\ref{TOV2}) is regular and its right-hand side is singular. Viceversa, if $\rho - \frac{M}{\frac{4\pi r^3}{3}}$ tends
to zero while $(r-2M)$ is finite the left-hand side of Eq. (\ref{TOV2}) is singular and its right-hand side is regular.
In both cases we come to a contradiction. Thus, the only regime possible is that, when both factors disappear at the same value
of $r = r_0$. The value of the energy density $\rho$ at this value of radius is
\begin{equation}
\rho(r_0) = \frac{3}{8\pi r_0^2}.
\label{crit}
\end{equation}

We shall call the surface with $r = r_0$ ``equator'' because after the crossing of this surface the radial variable
begin decreasing (cf. \cite{we-TOV}).
To describe the crossing of the equator it is convenient to introduce the trigonometric coordinate
\begin{equation}
r = r_0\sin \chi.
\label{trig}
\end{equation}
Now the TOV equations look like
\begin{equation}
\frac{d\rho}{d\chi} = \frac{2\cos\chi\rho\left(M-\frac43\pi r_0^3\sin^3\chi\rho\right)}{\sin\chi(r_0\sin\chi-2M)},
\label{TOV-trig}
\end{equation}
\begin{equation}
\frac{dM}{d\chi} = 4\pi r_0^3\sin^2\chi\cos\chi\rho.
\label{mass-trig}
\end{equation}
In the neighbourhood of the equator, where
\begin{equation}
\chi = \frac{\pi}{2} + \tilde{\chi},
\label{equat}
\end{equation}
 where $\tilde{\chi}$ is a small angle, the solution of Eqs. (\ref{TOV-trig}) and (\ref{mass-trig}) is
\begin{equation}
\rho = \frac{3}{8\pi r_0^2} + \rho_1 \tilde{\chi} + \frac{3}{8\pi r_0^2}\tilde{\chi}^2 + \cdots,
\label{equat1}
\end{equation}
where $\rho_1$ ia an arbitrary non-negative constant and
\begin{equation}
M = \frac{r_0}{2} -\frac34r_0\tilde{\chi}^2 + \cdots.
\label{mass-equat}
\end{equation}
Thus, we see that the regime of the equator crossing is defined by two constants $r_0$ and $\rho_1$, which
correspond to two initial conditions: $\rho(r_b)$ and $M(r_b)$.

Now we would like to study what happens with the geometry of the universe  after the crossing of the
equator.
It is convenient for this purpose to introduce a new variable
\begin{equation}
y \equiv \frac{1}{\sin\chi}.
\label{y-def}
\end{equation}
In terms of this variable the TOV equations can be rewritten as
\begin{equation}
\frac{d\rho}{dy} = -\frac{2\rho\left(My^3-\frac43\pi r_0^3\rho\right)}{y^3(r_0-2My)},
\label{TOV-y}
\end{equation}
\begin{equation}
\frac{dM}{dy} = -\frac{4\pi r_0^3 \rho}{y^4}.
\label{mass-y}
\end{equation}
From the formula (\ref{equat1}) it follows that after the crossing of the equator the energy density is growing.
That means that the factor $\left(My^3-\frac43r_0^3\rho\right)$ in the right-hand side of Eq. (\ref{TOV-y}) is decreasing becoming
negative. A simple calculation shows that $\frac{d(r_0-2My^3)}{dy}$ is equal to $2r_0$ at the equator crossing.
Thus, this factor is growing. The change of the behaviour of one of  two quantities, $\rho$ or $\left(My^3-\frac43r_0^3\rho\right)$, is possible only if one of them or the factor $(r_0-2My^3)$ would have passed through the zero value, but it is impossible, while $\rho$ is growing and  $\rho$ or $\left(My^3-\frac43r_0^3\rho\right)$ is decreasing, being negative.
That means that during the evolution in terms of change of the parameter $y$ in the interval $1  \leq y \leq \infty$
the factors $(r_0-2My^3)$ and $\left(My^3-\frac43r_0^3\rho\right)$ conserve their signs. Note, that the value $y = \infty$ corresponds
to the opposite pole of our spheroidal manifold, i.e. to the point where $\chi = \pi$.

In principle, one can imagine the situation, when the manifold cannot be continued until the pole
$\chi = \pi\  (y = \infty)$, because of the singularity arising at some finite value of $y = y_0$.
Such a singularity could arise if the the energy density is growing as
\begin{equation}
\rho(y) = \rho_2(y_0-y)^\alpha,\ \alpha < 0.
\label{sing}
\end{equation}
Then, the mass behaves as $(y_0-y)^{\alpha+1}$, i.e. it is much smaller than $\rho$.
In this case Eq. (\ref{TOV-y}) tells that $\frac{d\rho}{dy}$ in its left-hand side is proportional
to $\rho^2$ in its righ-hand side. It implies the equality $\alpha = -1$ which, in turn implies the
logaritmic divergence of the mass, which excludes the equality between  the left-hand side and
the right-hand sides of Eq. (\ref{TOV-y}).

Now, after having proved the impossibility of existence of the singularity at $y = Y_0 < \infty$, we should study
different possible regimes of approaching to the pole $y = +\infty$.
First, let us suppose that when $y \rightarrow \infty$ the energy density of gas $\rho$ tends to some finite value:
\begin{equation}
\rho = \rho_0 -\frac{\rho_2}{y^{\alpha}},
\label{pole}
\end{equation}
 where, $\rho_1, \rho_2$ and $\alpha$ some positive constants.
In this case from Eq. (\ref{mass-y}) follows that
\begin{equation}
M = M_0 + \frac{4\pi r_0^3}{3y^3}\rho_0,
\label{mass-pole}
\end{equation}
where $M_0$ is a negative constant.
Substituting expressions (\ref{pole}) and (\ref{mass-pole}) into Eq. (\ref{TOV-y}) we obtain the following relation:
\begin{equation}
\frac{\alpha\rho_2}{y^{\alpha+1}} = \frac{\rho_0}{y^2},
\label{pole1}
\end{equation}
which can be satisfied at $\alpha = 1, \rho_2 = \rho_0$.
Thus,
\begin{equation}
\rho = \rho_0 - \frac{\rho_0}{y}
\label{pole2}
\end{equation}
describes a possible regime of approximation to the pole $\chi = \pi$, this regime is characterized by two
constants : $\rho_0> 0$ and $M_0 < 0$. Trying to consider the regime when $M_0 = 0$, we come to the relation
$\alpha+1 = \alpha + 3$, which obviously cannot be satisfied.

Now, let us consider a regime when $\rho$ tends to infinity. The analysis similar to one done above shows that
such regime can exist if the energy density of the string gas diverges linearly
\begin{equation}
\rho = \rho_3 y,
\label{log}
\end{equation}
 where $\rho_3$ is a positive constant, while the mass tends to some negative value $M_2 < 0$.

Thus, we have find two families of geometries. One of them has a finite scalar curvature
at the south pole of the spheroidal manifold, while the second family characterized by the divergent energy density and, hence,
by the divergent scalar curvature. Both the families have a standard Schwarzschild-like singularity at the south pole but
with the negative value of the mass parameter.

It is impossible to establish analytically what is the relation between these two families and the parameters $r_0$
and $\rho_1$, describing the equator crossing. However, such a relation can be studied numerically.
We have studied the numerical solution of the couple of differential equations (\ref{TOV-y}) and (\ref{mass-y}) and have found 
that for all the positive values of the radius of equator $r_0$ and for all the positive values of the parameter $\rho_1$ the second regime, whose asymptotic behavior is given by Eq. (\ref{log}) is realized. Thus, only the second geometry with the energy density and scalar 
curvature diverging at the south pole does exist.    

\section{Motion around the spherically symmetric object surrounded by the string gas}
Let us consider now the rotation curves around spherical symmetric object surrounded by the string gas. 
The dependence of the velocity of rotating test mass around the spherically symmetric object in the presence of 
the string gas is given as usual by the formula
\begin{equation}
v = \sqrt{\frac{GM(r)}{r}}.
\label{velocity}
\end{equation}
Here $G$ is the Newton constant and $M(r)$ is the solution of equations (\ref{massTOV}) and (\ref{TOV1}).
We integrate numerically these equations with some initial conditions at the boundary $r=r_b$ of the spherically symmetric object.
The initial conditions satisfy the relations  (\ref{hor}) and (\ref{initial}). The numerical calculations in the vicinity of the 
spherically symmetric object shows that the rotation curve in the presence of the string gas is rather different from one in its absence.
The velocity is decreasing very slowly and after achieving some minimum value begin growing. However, this growing begins in the region 
where the approximation of the unique spherically symmetric object in the universe is already invalid and it is necessary to take 
into account the presence of other objects. In the figure the typical behaviour of the rotation curves in the presence and in the absence of the string gas is shown.
\begin{figure}[h]
\includegraphics[scale=1]{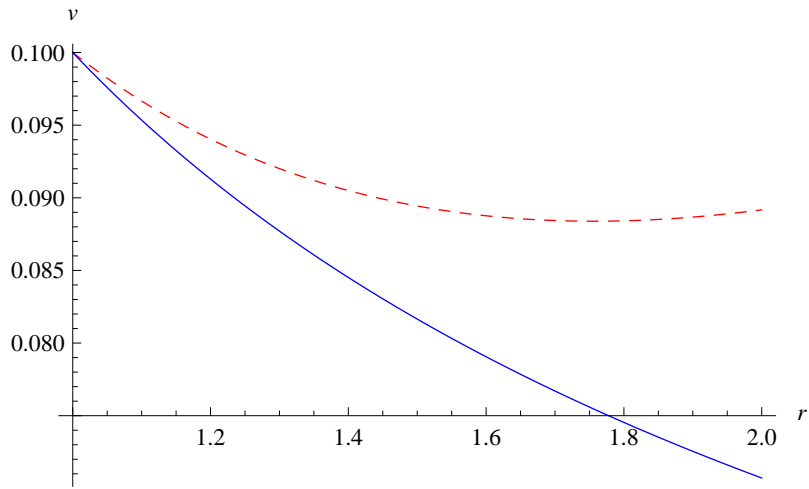}
\caption{The dependence of the velocity $v$ on the distance from the center of the spherical object. The red dashed line gives the rotation curve in the presence of the string gas, the blue line gives the rotation curve in its absence. The radius of the speherical object is chosen equal to 1. Its mass is equal to 0.01. The density of the string gas at the surface of the spherical object is chosen equal to 0.002 .}
\end{figure}
In principle one should also take into account the influence of the Archimedean force acting on the rotating object, which in our case modifies the formula (\ref{velocity}) as follows:
\begin{equation}
v = \sqrt{\frac{GM(r)}{r}-\frac{r}{3}\frac{d\rho(r)}{dr}\frac{1}{\rho_0}},
\label{velocity1}
\end{equation}
where $\rho_0$ is the density of the rotating object, which should much higher than the density of the 
string gas in the vicinity of the spherical object. The numerical calculations, however, show that the influence 
of the Archimedean term in the right-hand side Eq. (\ref{velocity1}) is negligibly small.
\section{Conclusions}
We have studied one of the simplest perfect fluids - the string gas with the equation of state (\ref{eq-of-state}) and have shown 
that unexpectingly it has rather interesting properties. In a Friedmann universe its presence can change the observable topology 
of the space, in the spherically symmetric spacetimes it produces rather bizzare geometries and in a way its influence on the 
rotation curves mimics the dark matter effects. 

\section*{Acknowledgements}
This work was partially supported by the RFBR  grant No 11-02-00643.

\end{document}